\documentclass[fleqn,usenatbib]{mnras}

\usepackage{newtxtext,newtxmath}

\usepackage[T1]{fontenc}
\usepackage{ae,aecompl}

\usepackage{graphicx}	
\usepackage{amsmath}	
\usepackage{amssymb}	

\usepackage[normalem]{ulem}

\title[QuasarNET]{QuasarNET: Human-level spectral classification and redshifting with Deep Neural Networks}

\author[N. G. Busca et al.]{
Nicolas G. Busca,$^{1}$\thanks{E-mail: nbusca@lpnhe.in2p3.fr} and
Christophe Balland,$^{1}$
\\
$^{1}$Sorbonne Universit\'e, Universit\'e Paris Diderot, CNRS/IN2P3,\\Laboratoire de Physique Nucl\'eaire et de Hautes Energies, LPNHE, 4 Place Jussieu, F-75252
Paris, France}

\date{Accepted XXX. Received YYY; in original form ZZZ}

\pubyear{2018}

\begin{document}
\label{firstpage}
\pagerange{\pageref{firstpage}--\pageref{lastpage}}
\maketitle

\begin{abstract}
We introduce QuasarNET, a deep convolutional neural network that performs classification and redshift estimation of astrophysical spectra with human-expert accuracy. We pose these two tasks as a \emph{feature detection} problem: presence or absence of spectral features  determines the class, and their wavelength determines the redshift, very much like human-experts proceed. When ran on BOSS data to identify quasars through their emission lines, QuasarNET defines a sample $99.51\pm0.03$\% pure and $99.52\pm0.03$\% complete, well above the requirements of many analyses using these data. 
QuasarNET significantly reduces the problem of line-confusion that induces catastrophic redshift failures to below 0.2\%. We also extend QuasarNET to classify spectra with broad absorption line (BAL) features, achieving an accuracy of $98.0\pm0.4$\% for recognizing BAL and $97.0\pm0.2$\% for rejecting non-BAL quasars. QuasarNET is trained on data of low signal-to-noise and medium resolution, typical of current and future astrophysical surveys, and could be easily applied to classify spectra from current and upcoming surveys such as eBOSS, DESI and 4MOST.

\end{abstract}




\section{Introduction}
\label{sec:intro}
The scientific exploitation of spectroscopic astrophysical data requires high-confidence spectral classification and a precise determination of the redshift. The volume of data that current and upcoming astrophysical surveys are collecting prohibits comprehensive visual inspection campaigns and the development of automatic methods achieving human-expert performance becomes essential \citep{Paris:2012iw,Paris:2016xdm}.

Standard automatic methods and human experts approach these tasks very differently. Automatic methods generically consist in comparing each spectrum with a data-base of spectral \emph{archetypes} \citep{Hutchinson:2016elj} to find the best matching class, or comparing the principal-component decompositions along different spectral classes \citep{Bolton:2012hz,Paris:2016xdm} to find a best fit solution. 

These methods perform significantly worse than human-experts, which motivated the inclusion of a comprehensive program of human-expert visual inspection of quasar targets to large astrophysical surveys such as the Baryon Oscillations Spectroscopic Survey survey (\citealt{Smee:2012wd,Gunn:2006tw, Eisenstein:2011sa,Dawson:2012va} and \cite{Paris:2016xdm} for the description of the visual inspection). Human-experts intervened to validate or correct  misclassifications (where the automatically-determined spectral class is incorrect) and catastrophic redshifts (where emission lines are classified incorrectly).

Human experts approach the task in a very different way from the automatic methods described above. They more or less immediately recognize spectral \emph{features} (emission lines, spectral breaks, absorptions, etc.) and use them for spectral classification. This classification and an \emph{eyeball} redshift based on the identified features can be used as a prior for a more precise automatic redshift fitter. Needless to say, the process of visually inspecting hundreds of thousands of spectra is tedious and requires a significant investment of human-expert time.

To limit the number of visually inspected quasar spectra in the extended Baryon Oscillation Spectroscopic Survey (eBOSS, \citealt{Dawson:2015wdb}), early data was used to develop a decision tree \citep{Paris:2017xme} based on the quality flags and the first five \emph{best-fit} solutions found by the automatic redshift fitter from \cite{Bolton:2012hz}. Spectra with bad quality flags or having inconsistent solutions among the top five best-fits would be automatically tagged as either non-quasar or as requiring a visual inspection. This procedure significantly reduces the fraction of spectra requiring a visual inspection down to less than about 5\% of the quasar targets, but a few tens of thousands still require it. The increase of data expected with upcoming large-scale-structure surveys such as DESI \citep{Aghamousa:2016zmz,Aghamousa:2016sne} turns visual inspection of a significant fraction of the spectra into a titanic effort.

The situation described above, where human-expert level performance is required and a large library of human-expert inspected data is available, constitutes a perfect setup for exploring the use of machine learning techniques. Recent efforts using such techniques in the context of the BOSS and eBOSS surveys, include quasar target selection \citep{Yeche:2009yh} and two \emph{damped-Lyman-$\alpha$} system detectors modeling the lyman-$\alpha$ forest as a gaussian process \citep{Garnett:2016awq} or with deep neural networks \citep{Parks:2017}.

In this article we introduce \emph{QuasarNET}, a convolutional neural network that achieves, as we demonstrate in this article, classification and redshift determination at human-expert levels of precision\footnote{Our code uses the open source python packages TensorFlow and Keras and is publicly available at https://github.com/ngbusca/QuasarNET.}. We train the network in a very similar way a human-expert would train another human: by ``teaching'' it to recognize spectral features and perform classification and redshifting in a subsequent step. During training, the network adapts its convolutional filters to recognize the portions of the spectrum where the features are, and ignore those where they are not, making the results more robust with respect to possible broad-band contaminations such as those due to imperfections in the flux extractions or atypical broad-band emission. In the context of recognizing quasar spectra, the features that we train the network to detect are quasar emission lines.

QuasarNET performs substantially better than the automatic approach described above, even when complemented with a significant fraction of visual inspections. It could eliminate the need for visual inspections in the current eBOSS and future surveys such as DESI \citep{Aghamousa:2016zmz,Aghamousa:2016sne} and 4MOST \citep{Roelof:2016}.

\section{Data Sample}
\label{sec:data_sample}
Our data sample is based on a large, publicly available data-base of spectra with human-expert classifications and redshift determinations \citep{Paris:2016xdm}, composed of 627,751 spectra of 546,856 objects from the Data Release 12 of the Baryon Oscillations Spectroscopic Survey (BOSS)  \citep{Alam:2015mbd} that were targeted as quasar candidates according to their color properties \citep{Ross:2011ky}.

Human-experts annotated this sample by classifying each spectrum with a class identifier, STAR (class id = 1), GALAXY (class id = 4), QSO (class id = 3) or QSO\_BAL (class id = 30, for quasars with Broad Absorption Lines), given in the CLASS\_PERSON column, a confidence, an integer between 1 (low) and 3 (high) given in the Z\_CONF\_PERSON column, and a redshift, given in the Z\_VI column of the \cite{Paris:2016xdm} \emph{superset} catalog\footnote{available at: https://data.sdss.org/sas/dr12/boss/qso/DR12Q/ Superset\_DR12Q.fits}. The classes may include a suffix ``\_?'' and have a confidence of 1 whenever the human-expert assigns a class with low confidence or a suffix ``\_Z\_?'' and have a confidence of 2 whenever the human-expert assigns a class with confidence but a redshift with low confidence. Firm classifications and redshifts are indicated with a confidence of 3. The Z\_VI redshifts consist mostly in a confirmation of the redshift determined by the automatic method from \citep{Bolton:2012hz} or an \emph{eyeball} correction whenever the human-expert found it necessary. To maximize the uniformity of the sample used in the present work, we exclude spectra of objects selected by ancillary programs.

Our final sample consists of 491,797 spectra (449,013 unique objects), of which 192,925 spectra (176,618 unique objects) are annotated as STAR, 14,966 spectra (14,203 unique objects) as GALAXY, 274,967 spectra (249,762 unique objects) as QSO spectra, while the remaining 8,939 spectra (8,430 unique objects) do not have a firm identification. Among the QSO spectra, 26,211 spectra (23,992 unique objects) were tagged as exhibiting \emph{broad absorption line} (BAL) features \citep{Weymann:1981dv,Weymann:1991zz}.

\section{Methods}
\label{sec:methods}
\begin{figure}
\includegraphics[width=\columnwidth]{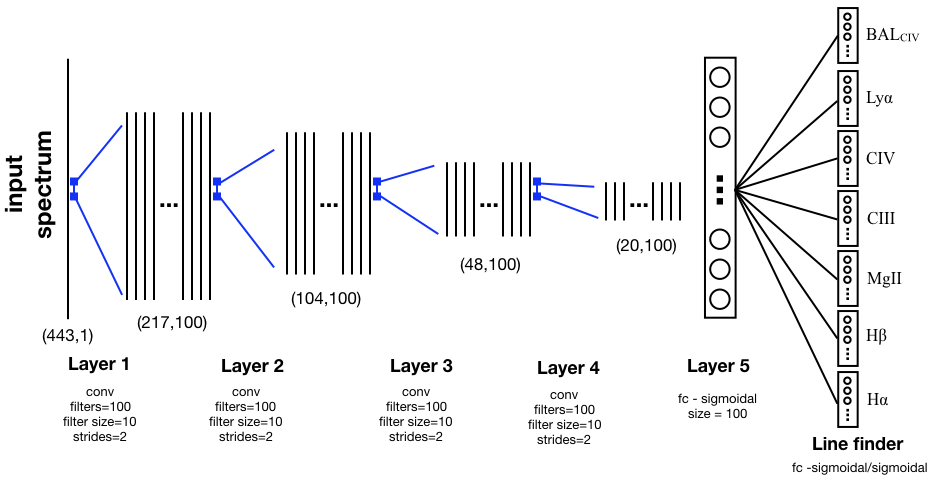}
\caption{Neural Network architecture. Layers 1-4 are convolutional layers of 100 filters of size 10 and strides of 2 and rectified linear unit (ReLU) activations. Layer 5 is a fully connected layer of size 100 and a sigmoid activation. Each line finder consists of 13 sigmoid units for the coarse-grained confidences and 13 sigmoid units for the fine-grained line position.}
\label{fig:architecture}
\end{figure}

We use the network architecture depicted in figure \ref{fig:architecture}. Raw spectra consist of nearly 4400 flux pixels that are equally spaced in log-wavelength between 360 nm and 1 $\mu$m. We downsample them to 443 pixels, also equally spaced in log-wavelength between the same limits. Large variations in the inputs to neural networks, in our case due to the distribution of quasar luminosity, are known to slow down learning. To reduce this problem we \emph{renormalize} spectra by subtracting to each flux its weighted mean and dividing the result by its weighted root-mean-square, using the inverse variances available with each spectrum as weights. These downsampled, renormalized spectra are fed to QuasarNET, successively reprocessed by four consecutive convolutional layers of 100 filters of size 10 pixels and strides of 2 with rectified linear unit (ReLU) activations, and finally encoded into a 100-dimensional vector by means of layer 5, a fully connected layer of 100 sigmoid-activation units. We found faster training performance by including a batch normalization \citep{Ioffe:2015} after each convolutional layer. This step renormalizes the outputs of each layer in a similar way as we renomalize the input spectra.

The output layer consists of \emph{line finders}, one for each emission line. Each line-finder is a \emph{feature detection} unit specialized for a given emission line. We implement this feature detection as a combination of a classification and a regression problem, inspired by recent techniques proposed for object-detection in images \citep{redmon2016yolo9000}. The input spectrum is divided into 13 equal-length intervals and the finder is trained to estimate, for each interval, a \emph{confidence} that the center of the line is contained within the interval via 13 sigmoid units. The interval with the maximum confidence provides a \emph{coarse-grained} position for the line. A fine-grained line position is obtained by another set of 13 sigmoid units that are trained to find the center of the line within each interval. We used 7 line-finders corresponding to Ly$\alpha$ (121.6 nm), CIV (154.9 nm), CIII (190.9 nm), MgII (279.6 nm), H$\beta$ (486.2 nm), H$\alpha$ (656.3 nm) and BAL\_CIV. 

We determined the parameters of the network by minimizing the loss function, $\cal L$:
\begin{align}
{\cal L} &= \sum_\ell\left[-\frac{1}{\sum_{i\alpha}Y_{i\ell\alpha}}\sum_{i\alpha}
Y_{i\ell\alpha}\ln\hat{Y}_{i\ell\alpha}\right.\nonumber\\
&-\frac{1}{\sum_{i\alpha}(1-Y_{i\ell\alpha})}\sum_{i\alpha}(1-Y_{i\ell\alpha})\ln(1-\hat{Y}_{i\ell\alpha}) \nonumber\\
&+\left.\frac{1}{\sum_{i\alpha}Y_{i\ell\alpha}}\sum_{i\alpha} Y_{i\ell\alpha}(X_{i\ell\alpha}-\hat{X}_{i\ell\alpha})^2\right] \label{eq:loss}
\end{align}
where the sum over $i$ runs over the sample of spectra, the sum over $\ell$ runs over the emission lines (from 1 to 7 in our case) and the sum over $\alpha$ runs over the wavelength intervals (13 in our case). $Y_{i\ell\alpha}$ is 1 for the $i$th spectrum if the center of the $\ell$th line is contained within the $\alpha$th interval and 0 otherwise. $Y_{i\ell\alpha}$ is zero for non-quasar spectra. $X_{i\ell\alpha}$ is, for the $i$th quasar spectrum, the offset of the $\ell$th line center with respect to the left edge of the $\alpha$th interval where the line lies. Quantities with a hat ($\hat{Y}_{i\ell\alpha}$, $\hat{X}_{i\ell\alpha}$) correspond to the predictions of the network for the corresponding quantity without a hat.

The first two terms in equation \ref{eq:loss}, similar to what is usually called a \emph{categorical cross-entropy} loss in the neural network community, is a binomial log-likelihood term with parameter \emph{p}=$\hat{Y}_{i\ell\alpha}$. We modify this standard loss in two ways. The standard categorical cross-entropy loss would require, first, that $\sum_\alpha Y_{i\ell\alpha}=1$ for all $i,\ell$, i.e., that all spectra have all emission lines within the spectrograph bounds, and, second, that the prefactors of the two terms be equal. The first condition is, of course, impossible to satisfy with our wavelength range limited to 3600\AA-10000\AA and the fact that many spectra do not even exhibit the emission lines listed above. The second condition would bias the estimates of the confidence, $\hat{Y}_{i\ell\alpha}$, towards zero since there are many more empty wavelength bins than wavelength bins with emission lines. Our choice of relative normalization ensures that both full and empty bins contribute equally to the loss. \cite{Ioffe:2015} tune the relative normalization of the terms to improve the performance on their validation sample, but we find that our simpler approach already leads to very good results for the task at hand (see \S \ref{sec:results}). 
The last term is a standard \emph{unweighted} least-squares loss. Its normalization relative to the other two terms was chosen such that for random values of  $\hat{Y}_{i\alpha\ell}$, resulting from the random initialization of the network, all three terms have roughly the same order of magnitude.

As mentioned above, we train QuasarNET to detect seven emission lines in the spectra of quasars: Ly$\alpha$ (121.6 nm), CIV (154.9 nm), CIII (190.9 nm), MgII (279.6 nm), H$\beta$ (486.2 nm) and H$\alpha$ (656.3 nm), plus a seventh CIV line with broad absorption features. These lines are chosen such that at least two of them are visible in the optical spectrograph of BOSS for quasar redshifts between 0 and 5.45 (the upper limit corresponds to the redshift at which the CIV line leaves the red end of the spectrograph). 

To minimize the loss in equation \ref{eq:loss} we use the \emph{adam} algorithm \citep{Kingma:2014}, which is based on stochastic gradient descent: the gradients of the loss are calculated over a small sample of data, a \emph{mini-batch} of 256 spectra in our case, and a gradient descent step is taken for every mini-batch. The algorithm adapts the learning rate according to the running average of the variance of the gradient, reducing it when this variance average increases. We used the AMSGrad option of the adam algorithm \citep{Reddi:2018on}, which prevents the learning rate from increasing, since it resulted in a more stable convergence. An \emph{epoch} is the set of mini-batches that covers the full training sample and over 1,500 gradient-descent steps are taken in every epoch. We train QuasarNET over 100 epochs.

\begin{figure}
\includegraphics[width=\columnwidth]{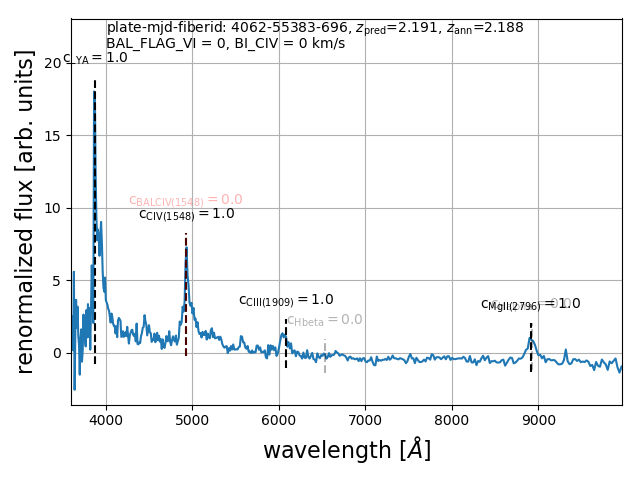}
\caption{An example of a high-redshift quasar spectrum from BOSS (identified by the plate id, the modified julian day (mjd) and the fiber id of the observation) annotated as a quasar and with high-confidence emission-line detections by QuasarNET. The broad absorption line annotations, BAL\_FLAG\_VI and BI\_CIV are also shown (see \S \ref{ssec:results_bal}). The flux has been renormalized as explained in the text. Also indicated are the positions of the lined as found by QuasarNET (black dashed lines) and the confidences. Low-confidence lines are grayed out.}
\label{fig:example_spectrum}
\end{figure}

The output of our network consists of a confidence for the detection of each emission line, a number between 0 and 1 that indicates the degree of certainty at which the network detected it, and the corresponding wavelength. Figure \ref{fig:example_spectrum} shows an example spectrum and the location and confidences of the lines detected by QuasarNET. 

We split the full sample into ten 80/20 training/validation random splits. For each realization, we train QuasarNET over the training sample and study the performance on the validation sample. To avoid correlations between these samples, we excluded from the validation sample spectra that correspond to targets also present in the training sample.

For each spectrum in the validation sample, QuasarNET predicts a confidence and a position for each emission line. A spectrum is likely to be that of a quasar if one or several emission lines are found with high confidence. 

The standard 24-cpu computer that we use takes about 12 minutes to train over a single epoch or about 20 hours for the full set of 100 epochs.

\section{Results}
\label{sec:results}
\subsection{The QuasarNET quasar sample}
\label{ssec:quasar_sample}

We define the QuasarNET quasar sample  as the set of spectra where the network finds a minimum number of lines, $\ell_\mathrm{min}$, with a confidence higher than a threshold value, $c_\mathrm{th}$, and use the wavelength of the most confident line to determine the redshift. This is a simple scheme that leads to very good performance. 

We characterize the quality of the sample by means of the \emph{purity} and \emph{completeness} measured on the the predicted quasar sample from the validation set. We define the purity as the ratio of the number of correctly predicted quasars to the total number of predicted quasars and the completeness as the ratio of the number of correctly predicted quasars to the total number of quasars. We consider that a quasar is correctly predicted if the absolute value of the velocity difference implied by the predicted redshift ($z_\mathrm{pred}$) and the annotated redshift ($z_\mathrm{ann}$), $\Delta v \equiv c(z_\mathrm{pred}-z_\mathrm{ann})/(1+z_\mathrm{ann})$ (where $c$ is the speed of light) is less than 6,000 km/s. Since there is some level of ambiguity between the QSO and GALAXY classes for galaxies with broad-line emission, we consider that spectra annotated as GALAXY are correctly identified by QuasarNET if $\Delta v$ satisfies the condition above. Our velocity requirement is a loose restriction on the redshift that discards $z_\mathrm{pred}$ values that would correspond to emission-line mis-identifications. For the purposes of measuring the completeness, we additionally require that the spectrum be annotated as a quasar by the human-expert. Both the purity and completeness are measured exclusively using spectra with secure annotations (Z\_CONF\_PERSON=3). 

We compare the quality of the sample predicted by QuasarNET to the automatic classification from BOSS \citep{Bolton:2012hz}, which we refer to as \emph{auto}, and the automatic classification complemented with visual inspections from eBOSS \citep{Paris:2017xme}, which we refer to as \emph{auto+VI}. The former is based on best fit classifications and redshifts using sets of PCA for three different classes (GALAXY, QSO, STAR) including a few subclasses for each class, and a quality flag, ZWARNING, that is set to zero for confident best fit models (see \citealt{Bolton:2012hz} for details). The latter is based on a decision tree that uses the first five best fit models from the automatic classifications and the ZWARNING flag. The decision tree either accepts the best fit result from the automatic procedure or tags the spectrum as requiring visual inspection. We assign to all spectra requiring visual inspection the annotated class and redshift. For the \emph{auto} (\emph{auto+VI}) quasar sample we obtain a purity of 99.14\% (99.17\%) and a completeness of 94.98\% (99.26\%). These numbers differ slightly from those stated in \cite{Paris:2017xme} because our data sample has a different redshift distribution and we include a quality condition on the redshift to consider a classification as correct. 

\begin{figure}
\includegraphics[width=\columnwidth]{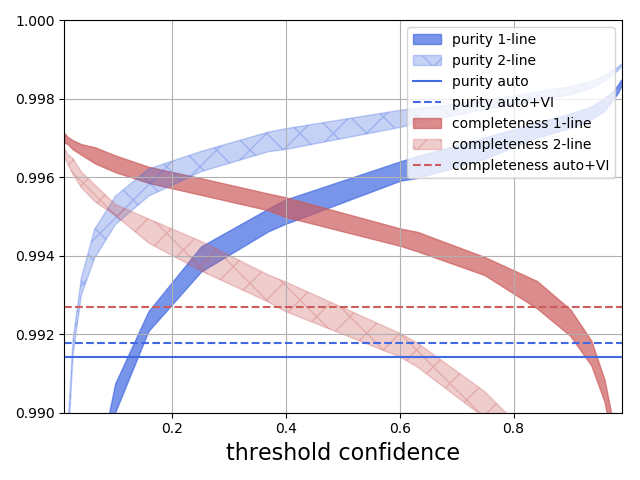}
\caption{Purity (blue) and completeness (red) of the predicted quasar sample as a function of threshold confidence for one 1 line (solid) or two lines (hatched). Bands represent the root-mean-square calculated over ten random 80/20 training/validation data splits. The horizontal lines correspond to the purity (blue) and completeness (red) for the automatic procedure from BOSS (solid) and the automatic procedure improved by visual inspections from eBOSS (dashed).}
\label{fig:pur_comp}
\end{figure}

Figure \ref{fig:pur_comp} shows the purity and completeness as a function of the threshold confidence for 1-line and 2-line detections. The choice of threshold confidence and minimum number or lines depends on the desired levels of purity and completeness. For definiteness, we adopt a nominal threshold confidence ($c_\mathrm{th}$)  of 0.4 and a single-line detection,  where the purity and completeness are roughly equal. For this threshold value we obtain a purity of to $99.51\pm0.03$\% and a completeness of $99.52\pm0.03$\% (the uncertainties indicate the root-mean-square over the training/validation splits). We would obtain similar purity and completeness for two-line detections and a lower threshold confidence, near 0.1. For three-line detections, not shown in the figure, we find a higher purity, above 99.9\% for $c_\mathrm{th}=0.6$, at the expense of a significantly degraded completeness, below 45\%.

We define a \emph{misidentified} quasar sample as composed of spectra classified as quasars by QuasarNET (according to our nominal definition above) but annotated as STAR ($0.27\pm0.02$\%), annotated as GALAXY ($0.19\pm0.03$\%), or where the predicted redshift significantly differed from the annotated redshift ($0.14\pm0.01$\%) and visually inspected it. Most spectra classified as STAR in this sample exhibit emission-line-like features that are most likely noise but still triggered QuasarNET. However, we still found a few of them (less than about 10\%) where the emission-lines are clearly present. A few example spectra in this class are shown in appendix \ref{app:star_as_qso}.

For spectra classified as GALAXY in the misidentified sample, we found that over 80\% had a predicted redshift that agreed with the annotated redshift. We visually inspected a few of such spectra and found that they were galaxy-like spectra exhibiting broad emission lines correctly identified by QuasarNET. A few example spectra are shown in appendix \ref{app:galaxy_as_qso}.

We also visually inspected the sample of spectra annotated as QSO where QuasarNET detects emission lines but the predicted redshift significantly differs from the annotated redshift according to the criterion mentioned above. Most of the detections from QuasarNET in this sample correspond to misidentified emission lines or noise features, but we found a significant fraction, nearly 30\% of the spectra in the sample, where it is the annotated redshifts that are most likely wrong. In section \ref{app:quasar_cata_z} we show a few examples of spectra illustrating these two cases.

We also define a \emph{missed} quasar sample as composed of spectra with no emission lines detected above our nominal threshold confidence, but annotated as QSO ($0.33\pm0.02$\% of the number of spectra annotated as QSO) and spectra with emission lines detected but a predicted redshift that does not satisfy our redshift quality condition ($0.14\pm0.01$\% of the number of spectra annotated as QSO). 

We now turn to the sample of spectra without confident annotations, i.e., with Z\_CONF\_PERSON less than three, which represents about 2\% of the spectra. These spectra are mostly of too low signal-to-noise to be confidently classified by the human-expert. More than half of those spectra have tentative redshift annotations. In this 2\% sample, QuasarNET classifies 20\% of spectra as quasars with a redshift that agrees with the tentative redshift in over 60\% of the cases.

\begin{figure}
\includegraphics[width=\columnwidth]{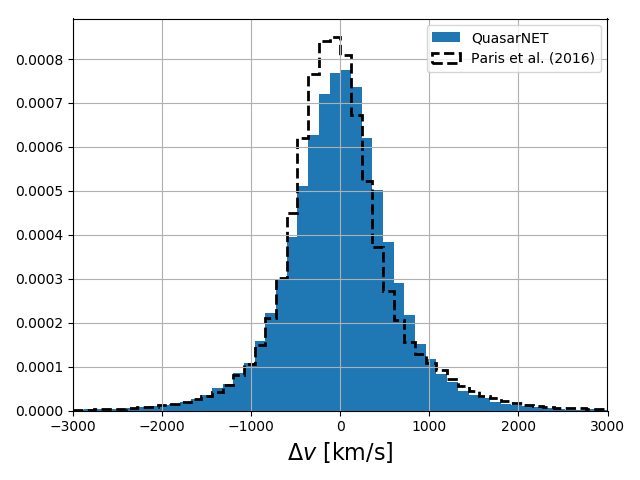}
\caption{Velocity difference implied by the predicted and annotated redshifts (blue), and Z\_VI and Z\_PCA redshifts from \citep{Paris:2016xdm} (black dashed line).}
\label{fig:dv_comp}
\end{figure}

For each spectrum we determine QuasarNET's redshift prediction from the most confident emission line. In figure \ref{fig:dv_comp} we compare the redshift of QuasarNET to the annotated redshift for quasar spectra found by the network in terms of $\Delta v$. On average, QuasarNET redshifts are redshifted by 8 km/s and have a dispersion of 664 km/s around the annotated redshift. These values are comparable to those found in \cite{Paris:2016xdm} when comparing different redshift estimates from either different fitters or based on different emission lines. For reference, the dashed black line in figure \ref{fig:dv_comp} shows the velocity difference between the redshift Z\_VI and Z\_PCA from the \cite{Paris:2016xdm} catalog.

\subsection{Classification of spectra with Broad Absorption Line features.}
\label{ssec:results_bal}

The problem of detecting quasars with a broad absorption line (BAL) feature \citep{Weymann:1981dv} is slightly more challenging than that of detecting emission lines. Quasars with a BAL feature exhibit a characteristic absorption trough blue-wards of the emission lines required to be sufficiently detached from the center of the line (at least 3000 km/s) and sufficiently wide (at least 2000 km/s) \citep{Weymann:1991zz}. The balnicity-index (BI) measures the \emph{strength} of the BAL feature as the equivalent width of the absorption trough, in excess of the minimum requirement of 2000 km/s. The main difficulty in determining whether a spectrum has a BAL feature, both visually and automatically, resides in the determination of the level of the unabsorbed continuum. For visual classifications, some degree of subjectivity also comes into play for limiting cases or low signal-to-noise ratios.

The sample annotated in \citet{Paris:2016xdm} includes both a visual BAL flag (BAL\_FLAG\_VI) and an automated measurement of the BI for the CIV line (BI\_CIV). Out of the 222,859 quasar spectra with redshift higher than 1.6 (where the CIV emission line enters the BOSS spectrograph), there are 27,815 that were tagged visually as BAL and 16,260 with BI\_CIV$>$0. The purity and completeness of either sample is not precisely known \citep{Paris:2012iw}.

From the perspective of feature detection, the problem of finding a CIV emission line with broad absorptions is equivalent to the problem of finding any of the other emissions lines, and we treat it as such by adding a BAL\_CIV line to the list of lines that the network learns to identify. However, to increase the purity of both positive (presence) and negative (absence) examples, we set up the BAL\_CIV line finder to assign a weight of zero to spectra either tagged as BAL with BI\_CIV=0 or not tagged as BAL with BI\_CIV$>$0. 

In the absence of a robust CIV BAL annotation, we characterize the performance of QuasarNET's CIV BAL finder by defining a \emph{pseudo-purity} as the ratio of the number of predicted CIV BAL spectra that also are visually tagged as BAL or have BI\_CIV $>0$ to the total number of predicted CIV BAL spectra. Similarly, we define a \emph{pseudo-completeness} as the ratio of the number of predicted CIV BAL spectra that are visually tagged as BAL \emph{and} have BI\_CIV $>$ 0 and the total number of spectra tagged as BAL that have BI\_CIV $>$ 0.

\begin{figure}
\includegraphics[width=\columnwidth]{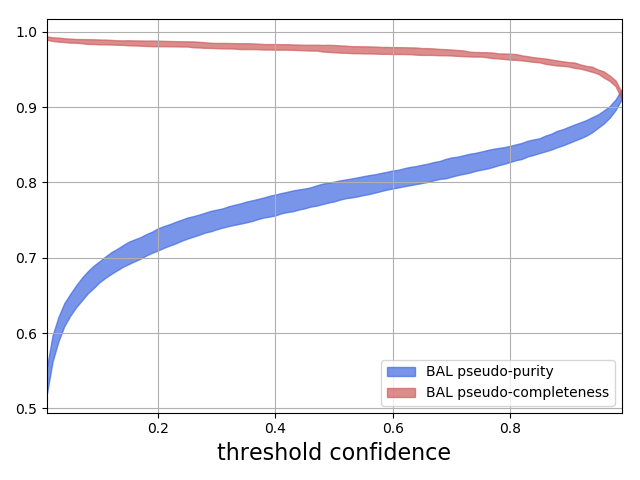}
\caption{\emph{pseudo-purity} (blue) and \emph{pseudo-completeness} (red) for the detection of quasar spectra exhibiting a BAL feature bluewards of the CIV emission line as a function of threshold confidence.}
\label{fig:bal_pur_comp}
\end{figure}

Figure \ref{fig:bal_pur_comp} shows the pseudo-purity and the pseudo-completeness as a function of the threshold confidence for a CIV BAL detection. Many analyses, like baryon acoustic oscillations based on lyman-alpha forest data \citep{Bautista:2017zgn,Bourboux:2017cbm,GontchoAGontcho:2017hru,Blomqvist:2018efj}, prefer high completeness to high purity, since the fraction of BAL spectra is already small (at the 10\% level). For definiteness, we adopt a nominal threshold confidence for CIV BAL detection at 0.4, which implies a completeness of $98.0\pm0.4$\% and a purity of $77\pm1$\%.

Another way to characterize the efficiency of the BAL CIV detection is via the accuracy for recognizing BAL quasars (equivalent to the pseudo-completeness in our case) and that for recognizing non-BAL quasars. QuasarNET identifies as BAL $98.0\pm0.4$\% of spectra visually annotated as BAL with BI\_CIV>0 and identifies as non-BAL $97.0\pm0.2$\% of spectra annotated as non-BAL with BI\_CIV=0.

We visually inspected the \emph{misidentified-BAL} quasar sample, composed of spectra satisfying our BAL criterion above, but annotated with zero balnicity index and zero visual BAL flag.  Most of the spectra in this sample exhibit absorption lines that are either too narrow or too close to the emission line to be considered BAL. 

\begin{figure}
\includegraphics[width=\columnwidth]{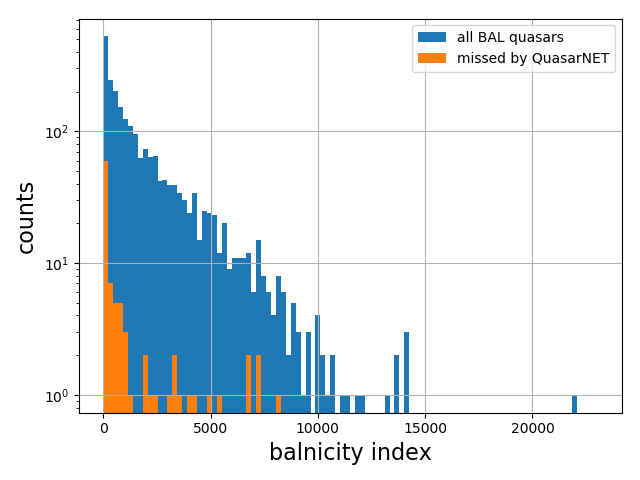}
\caption{Histogram of the CIV balnicity index in the QuasarNET quasar sample (blue) and in the QuasarNET missed-BAL quasar sample (orange). Only strictly positive values of the balnicity index are included in the histogram.}
\label{fig:bal_bi_dist}
\end{figure}

The missed-BAL quasar sample from QuasarNET is composed of spectra with confident emission-line detections but with BAL-CIV confidence below the nominal threshold. To further characterize this sample, we show in figure \ref{fig:bal_bi_dist} the distribution of the balnicity index BI\_CIV from \cite{Paris:2016xdm} of all spectra with BI\_CIV>0 and those in the missed-BAL quasar sample. While the median of BI\_CIV values in the full sample is nearly 936 km/s, that for the missed-BAL quasar sample is roughly 122 km/s. This implies that most spectra in the missed-BAL sample have a much narrower BAL feature than the typical BAL spectrum.

We studied the quality of the redshifts predicted by QuasarNET in its BAL-quasar sample. The fraction of predicted redshifts that are significantly different from the annotated redshift is very close to that in the full sample, at the level of 0.2\%. We do note a higher dispersion of 797 km/s (cf. 661 km/s for the full sample), although it is likely that the annotated redshift is noisier for BAL quasars as well.

\subsection{Performance vs. redshift}

In order to study the performance of QuasarNET as a function of redshift we organized spectra in the validation sample according to their redshifts in 11 redshift bins between 0 and 5.5. In this section we only use spectra with confident annotations (Z\_CONF\_PERSON=3).

To measure the purity as a function of redshift, we binned spectra according to the predicted redshifts from QuasarNET, the auto or auto+VI procedures. The purity in each bin is the ratio of the number of correctly predicted quasar spectra in the bin to the total number of predicted quasars in the bin. 

To measure the completeness as a function of redshift we only consider spectra annotated as QSO and bin them according to their annotated redshifts. The completeness in each bin is the ratio of the number of correctly predicted quasar spectra in the bin to the total number of spectra in the bin.

\begin{figure}
\includegraphics[width=\columnwidth]{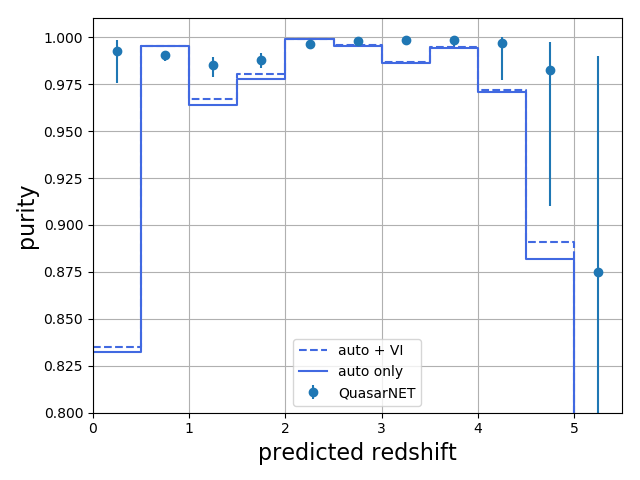}
\includegraphics[width=\columnwidth]{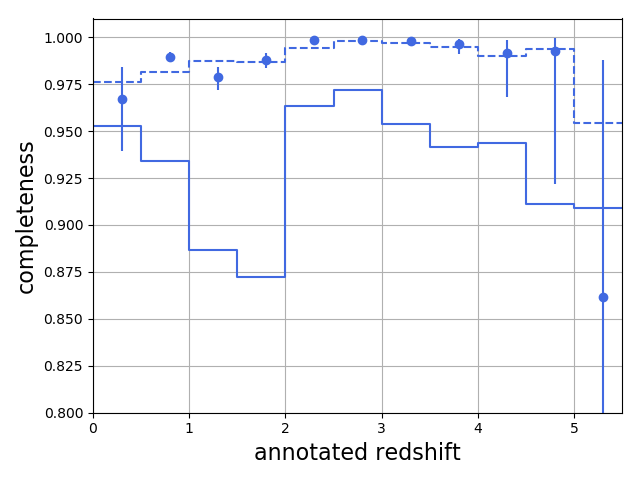}
\caption{Purity (top) and completeness (bottom) in each redshift bin. Errorbars indicate bayesian 95\% credible intervals.}
\label{fig:pur_com_vs_z}
\end{figure}

The purity and completeness as a function of redshift are shown in figure \ref{fig:pur_com_vs_z}. 

The purity of the QuasarNET quasar sample (top panel in figure \ref{fig:pur_com_vs_z}) is consistently better than that of the auto and auto+VI procedures, specially at very high redshifts where both the auto and auto+VI procedures are significantly degraded. This drop is a known issue due low redshift quasar spectra that contaminate the high redshift bin whenever the MgII line is mistaken by Lyman-$\alpha$ emission by the redshift fitter. The purity of the QuasarNET sample also drops at high redshift, but the drop is of a different nature. We find, instead, that the contaminants are high-redshift quasars of degraded predicted-redshift precision, at the 10,000 km/s level. The purity of the auto and auto+VI samples is also particularly low in the first redshift bin. Most of the offending spectra in this bin are annotated as STAR or GALAXY with a redshift significantly different from the annotated redshift (GALAXY spectra with the correct redshift are considered correct identifications). 

The completeness of the QuasarNET sample (bottom panel in figure \ref{fig:pur_com_vs_z}) is significantly better than that of the auto sample and at the level of that of the auto+VI sample except in the lowest and the highest redshift bin. In the lowest redshift bin, most of the quasar spectra missed by QuasarNET are below the detection threshold while a smaller fraction have a predicted redshift significantly different from the annotated redshift. In the highest redshift bin, the lost spectra have a redshift such that the CIV emission line is close to the red edge of the spectrograph.

\section{Discussion}
\label{sec:discussion}
In this article, we describe QuasarNET, a deep convolutional neural network trained to detect quasar emission lines in optical, medium resolution, medium signal-to-noise spectra. QuasarNET is trained using the existing large sample of nearly 500,000 quasar-target spectra from BOSS \citep{Paris:2016xdm}, which was comprehensively visually inspected and annotated by human-experts who verified and corrected spectral classifications and redshifts. QuasarNET also detects spectra exhibiting broad absorption line (BAL) bluewards of the CIV emission line.

By comparing the predictions of the network on a validation set, independent of the sample use for training, we demonstrate that the quasar sample defined by QuasarNET is of very high purity, over 99.5\%, and completeness, over 99.5\%, well above quality requirements set by large-scale surveys such as BOSS \citep{Dawson:2012va}, eBOSS \citep{Dawson:2015wdb} and DESI \citep{Aghamousa:2016zmz,Aghamousa:2016sne}. The quality of the QuasarNET sample is significantly better than that obtained using methods that use physical models of the spectra \citep{Bolton:2012hz}, and at the level of what is obtained when those methods are complemented with a significant fraction of human-expert visual inspection for uncertain spectra \citep{Paris:2017xme}. 

The network also predicts redshifts based on the wavelength of the detected emission lines. When compared to the annotated redshift, the predicted redshift implies a typical velocity difference, $\Delta v$, of 661 km/s. The fraction of catastrophic redshifts (defined as $|\Delta v|>6000$ km/s) is below 0.2\%. If better redshift precision is required, the redshift provided by QuasarNET could be used as a prior for more precise methods using physical models of the spectra. We note, however, that QuasarNET redshifts are uncertain at the level of what is obtained with different principal component sets, as illustrated in figure \ref{fig:dv_comp}.

QuasarNET could be used to significantly reduce the fraction of spectra requiring visual inspection in eBOSS \citep{Dawson:2015wdb}. For instance, using two-line detections to automatically classify quasars, and visually inspecting single-line detections. This scheme would maintain the completeness of the single-line detections (at the 99.5\% level) while increasing the purity above that of two-line detections (above 99.7\%) with only about 0.2\% of visual inspections.

Training QuasarNET is relatively inexpensive, about 600 cpu-hours for our 400,000 spectra training sample, and predictions take about 750 ns per spectrum on our standard 24-cpu unit. 

The data in the training sample are of a characteristic signal-to-noise of 1 in 1-\AA~ wavelength pixels and were taken using a spectrograph of resolution near 2000, values typical of upcoming surveys such as DESI \citep{Aghamousa:2016zmz,Aghamousa:2016sne} and 4MOST \citep{Roelof:2016}. To the extent that the objects targeted by those surveys are similar to those observed by BOSS, we expect QuasarNET to perform similarly well on their future data, even using the current training sample based on BOSS data, or complemented with a fraction of simulated spectra from those surveys.

\section*{Acknowledgements}
The authors would like to thank Julien Guy, Stephen Bailey and James Rich for useful discussion and comments. Funding for SDSS-III has been provided by the Alfred P. Sloan Foundation, the Participating Institutions, the National Science Foundation, and the U.S. Department of Energy Office of Science. The SDSS-III web site is http://www.sdss3.org/.

SDSS-III is managed by the Astrophysical Research Consortium for the Participating Institutions of the SDSS-III Collaboration including the University of Arizona, the Brazilian Participation Group, Brookhaven National Laboratory, Carnegie Mellon University, University of Florida, the French Participation Group, the German Participation Group, Harvard University, the Instituto de Astrofisica de Canarias, the Michigan State/Notre Dame/JINA Participation Group, Johns Hopkins University, Lawrence Berkeley National Laboratory, Max Planck Institute for Astrophysics, Max Planck Institute for Extraterrestrial Physics, New Mexico State University, New York University, Ohio State University, Pennsylvania State University, University of Portsmouth, Princeton University, the Spanish Participation Group, University of Tokyo, University of Utah, Vanderbilt University, University of Virginia, University of Washington, and Yale University.




\bibliographystyle{mnras}
\bibliography{biblio} 



\appendix

\section{Misclassification examples}
\subsection{STAR as quasar}
\label{app:star_as_qso}
\begin{figure*}
\includegraphics[width=\columnwidth]{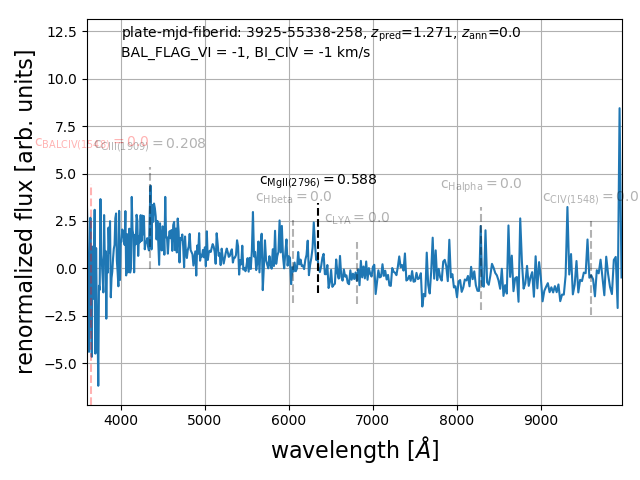}
\includegraphics[width=\columnwidth]{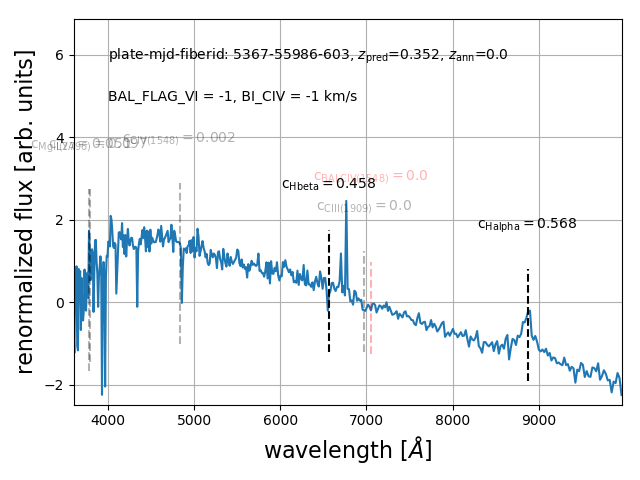}
\includegraphics[width=\columnwidth]{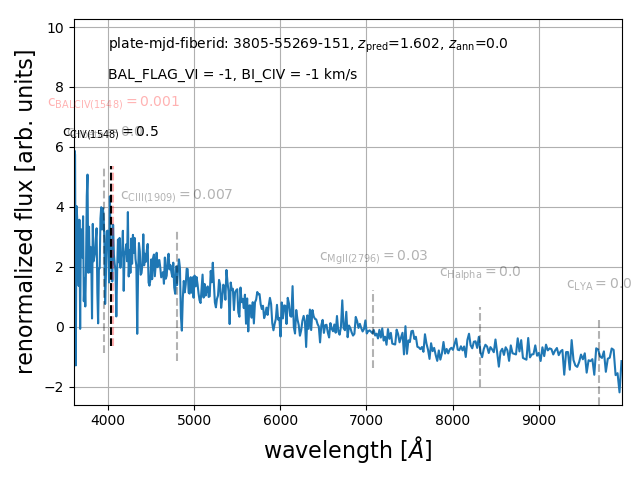}
\includegraphics[width=\columnwidth]{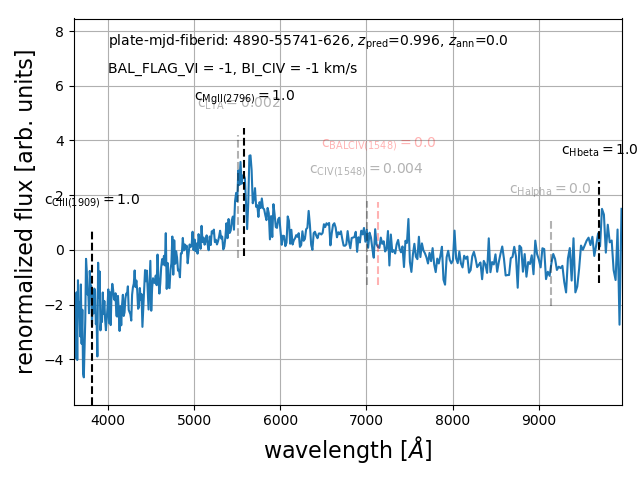}
\includegraphics[width=\columnwidth]{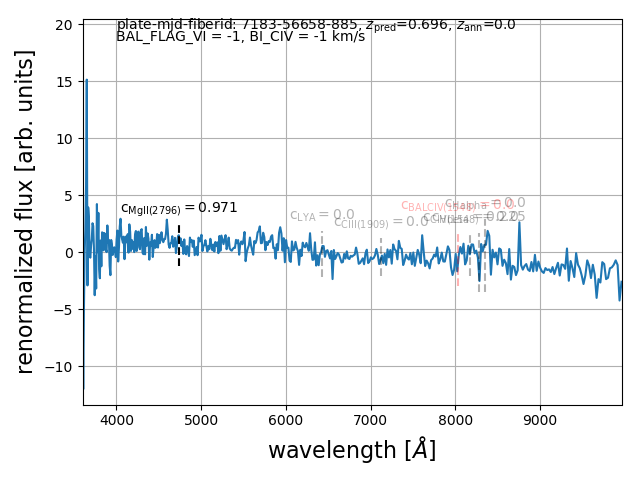}
\includegraphics[width=\columnwidth]{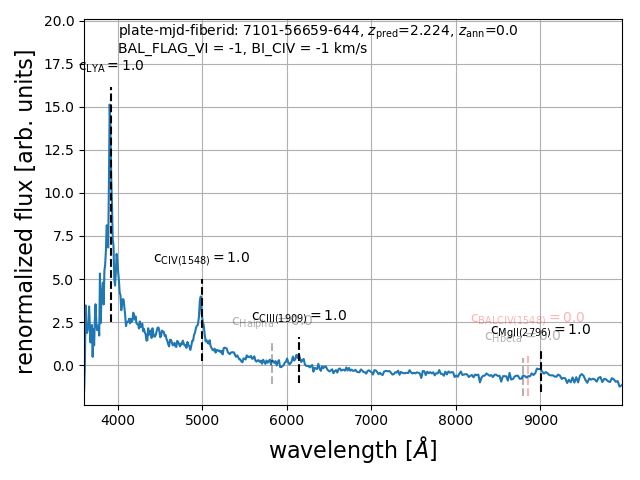}
\caption{Examples of spectra annotated as STAR but classified as quasar by QuasarNET. The left column shows examples where the emission lines detected by QuasarNET are most likely noise. The right column shows examples where the emission lines found by QuasarNET are clear features in the spectra.}
\label{fig:example_star_as_qso}
\end{figure*}

Figure \ref{fig:example_star_as_qso} shows examples of spectra annotated as STAR where QuasarNET detects emission lines. Examples where the detections by QuasarNET are most likely noise (left column) and where the detections correspond to visible features (right column) are shown.

\subsection{GALAXY as quasar}
\label{app:galaxy_as_qso}
\begin{figure*}
\includegraphics[width=\columnwidth]{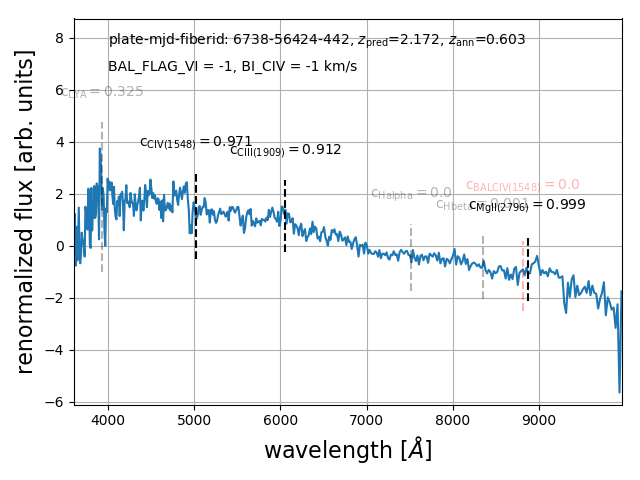}
\includegraphics[width=\columnwidth]{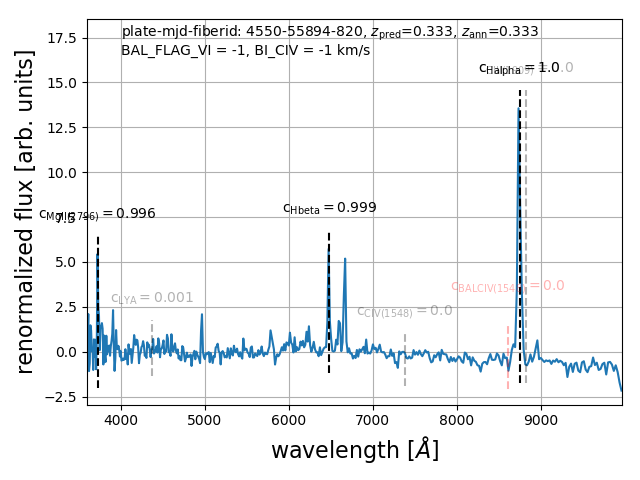}
\includegraphics[width=\columnwidth]{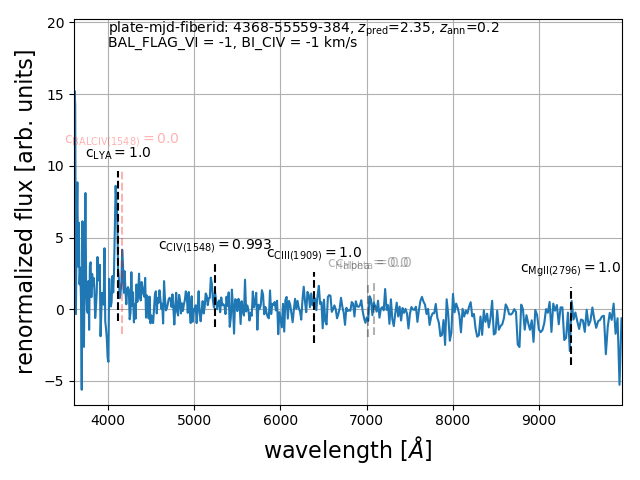}
\includegraphics[width=\columnwidth]{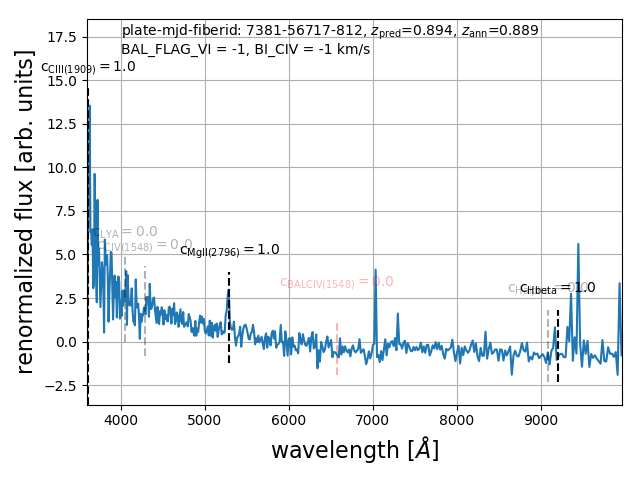}
\includegraphics[width=\columnwidth]{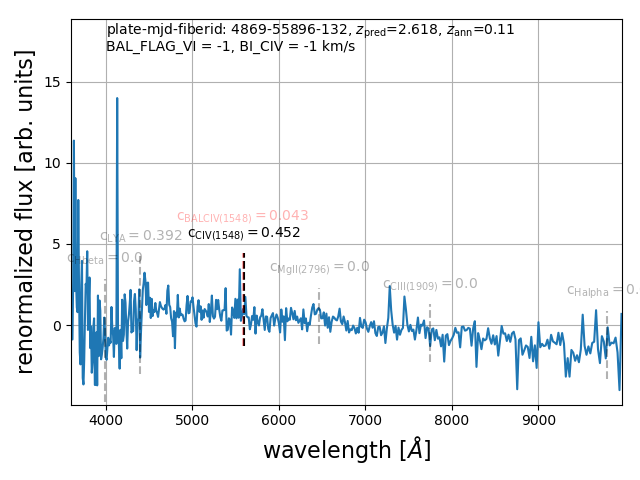}
\includegraphics[width=\columnwidth]{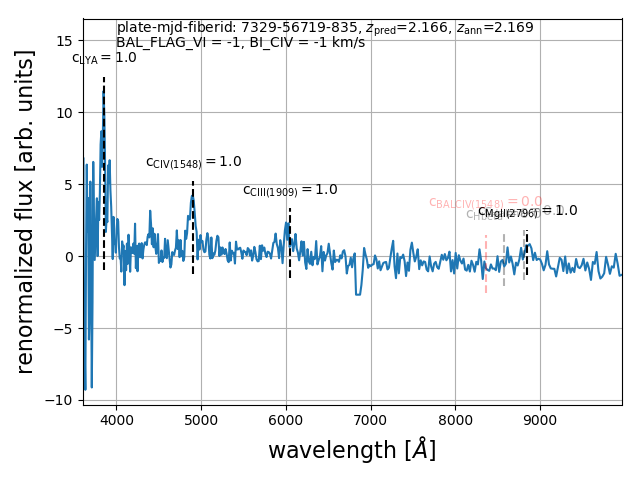}
\caption{Examples of spectra annotated as GALAXY where QuasarNET detects broad emission lines. The left column shows examples where the emission lines detected by QuasarNET are most likely noise. The right column shows examples where the emission lines found by QuasarNET are clear features in the spectra.}
\label{fig:example_gal_as_qso}
\end{figure*}

Figure \ref{fig:example_gal_as_qso} shows examples of spectra annotated as GALAXY where QuasarNET detects emission lines. Both examples where the detections by QuasarNET are most likely noise (left column) and where the detections correspond to visible features (right column) are shown.

\subsection{Quasars with discrepant predicted and annotated redshifts}
\label{app:quasar_cata_z}

\begin{figure*}
\includegraphics[width=\columnwidth]{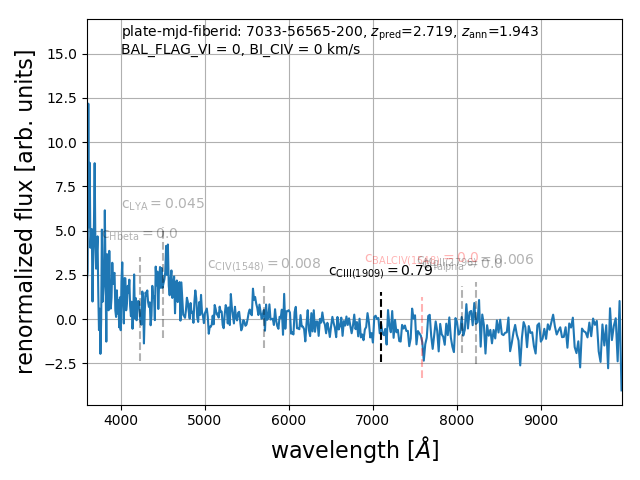}
\includegraphics[width=\columnwidth]{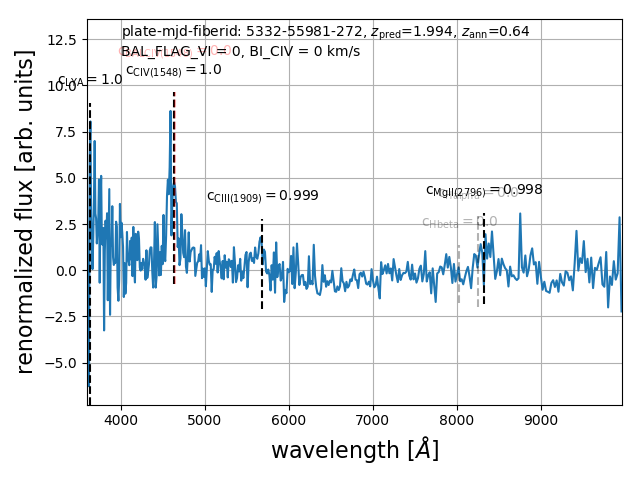}
\includegraphics[width=\columnwidth]{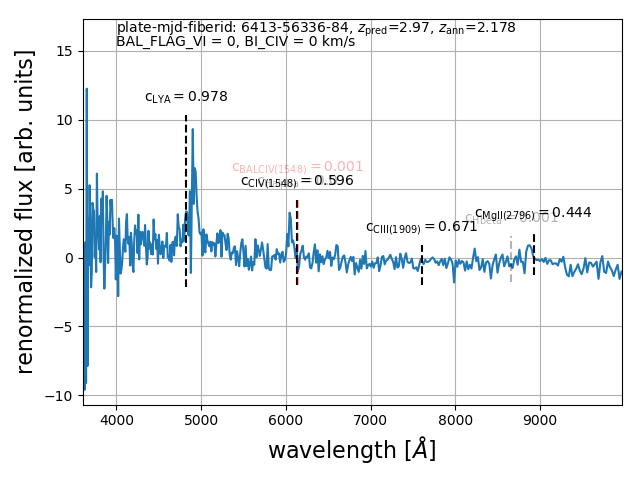}
\includegraphics[width=\columnwidth]{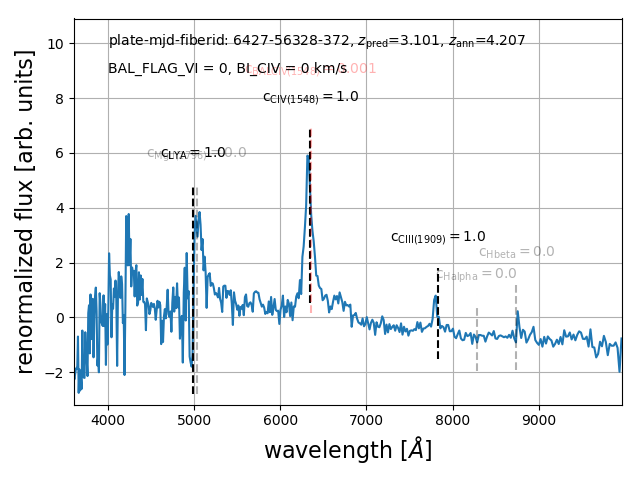}
\includegraphics[width=\columnwidth]{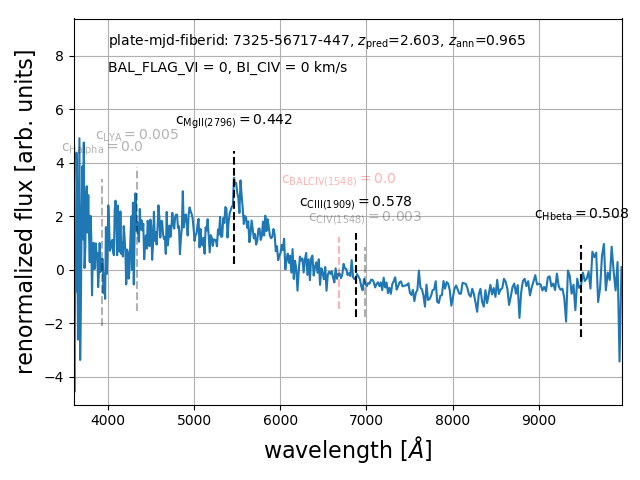}
\includegraphics[width=\columnwidth]{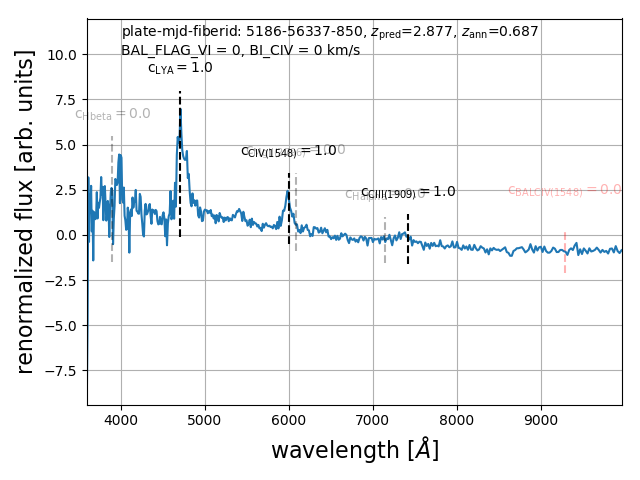}
\caption{Examples of spectra annotated as QSO where the predicted and the annotated redshifts are significantly different. The left column shows examples where the predicted redshift if most likely wrong. The right column shows examples where the the annotated redshift is most likely wrong.}
\label{fig:example_qso_cata_z}
\end{figure*}

Figure \ref{fig:example_qso_cata_z} shows example spectra annotated as QSO, where QuasarNET finds significant emission lines, but the annotated redshift significantly differs from the predicted redshift.

\subsection{Examples of spectra with broad BAL features}
\label{app:quasar_bal}

\begin{figure*}
\includegraphics[width=\columnwidth]{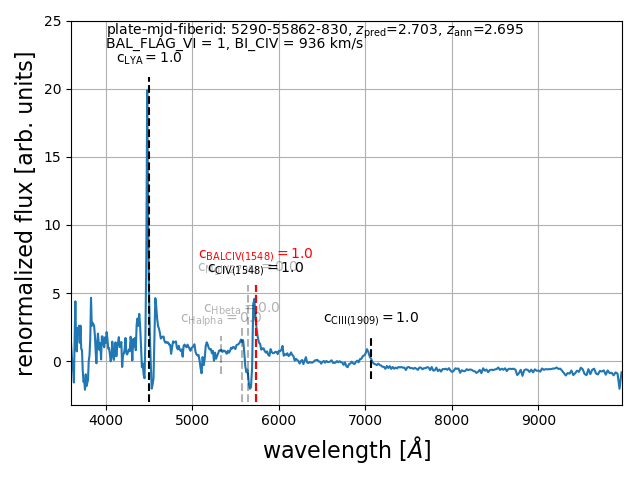}
\includegraphics[width=\columnwidth]{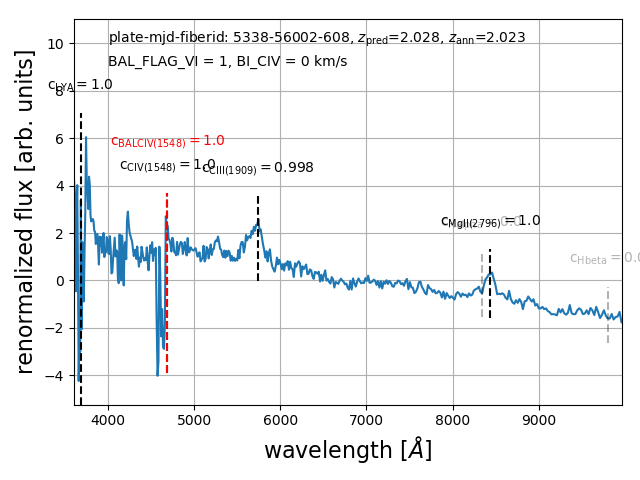}
\includegraphics[width=\columnwidth]{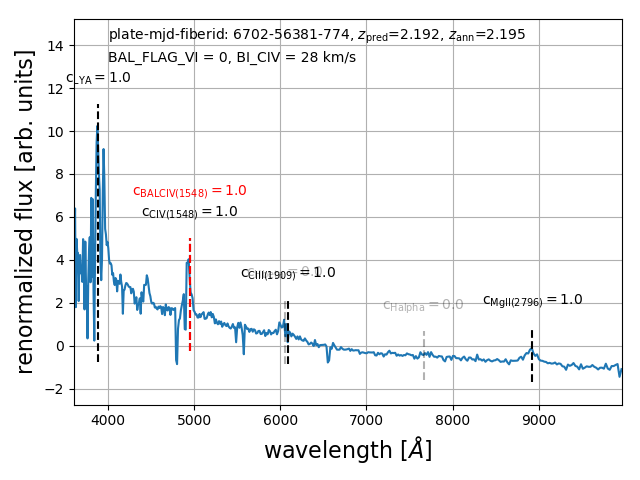}
\includegraphics[width=\columnwidth]{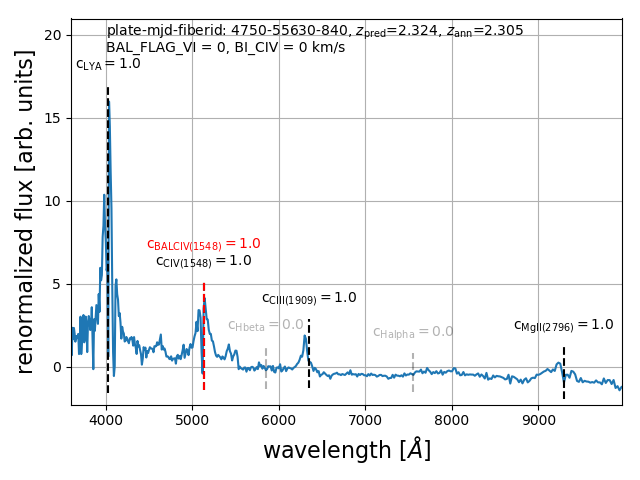}
\includegraphics[width=\columnwidth]{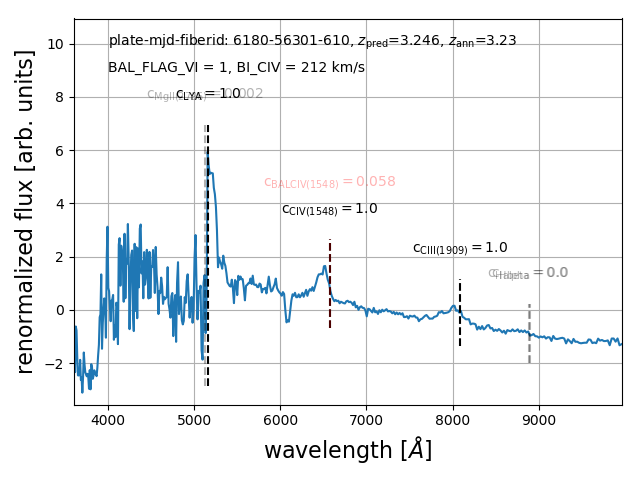}
\includegraphics[width=\columnwidth]{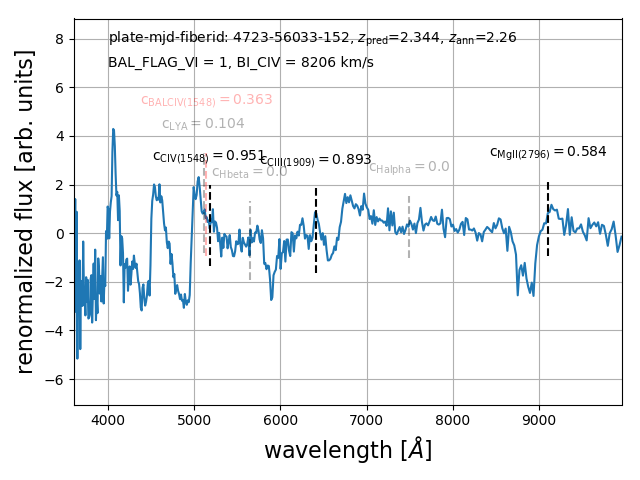}
\caption{Examples of CIV BAL spectra. The first three panels (from top to bottom and left to right) show examples spectra where QuasarNET detects a BAL feature with BAL\_FLAG\_VI=1 or BI\_CIV>0. The fourth panel is a typical false positive. The two bottom panels show spectra annotated as BAL but missed by QuasarNET. The bottom-left shows an example of a typical missed-BAL and the bottom right the most extreme missed-BAL in our sample.}
\label{fig:example_bal}
\end{figure*}

Figure \ref{fig:example_bal} shows example spectra where QuasarNET detects a CIV BAL feature (top four panels) or it does not (bottom two panels) but the spectra are annotated as BAL. 

\bsp	
\label{lastpage}
\end{document}